# AI-Enhanced Multi-Dimensional Measurement of Technological Convergence through Heterogeneous Graph and Semantic Learning

Siming Deng[1,2,4], Runsong Jia[2], Chunjuan Luan[3,4*], Mengjia Wu[2], Yi Zhang[2]

[1] School of Economics and Management, Dalian University of Technology, Dalian, 116023, China.
[2] Australian Artificial Intelligence Institute, Faculty of Engineering and Information Technology, University of Technology Sydney, Sydney, NSW 2007, Australia.
[3] School of Business, Dalian University of Technology, Panjin, 124221, China.
[4] School of Public Administration and Policy, Dalian University of Technology, Dalian, 116023, China.

**Abstract:** Technological convergence refers to the phenomenon where boundaries between technological areas and disciplines are increasingly blurred. It enables the integration of previously distinct domains and has become a mainstream trend in today's innovation process. However, accurately measuring technological convergence remains a persistent challenge due to its inherently multidimensional and evolving nature. This study designs an AI-enhanced Technological Convergence Index (TCI) that comprehensively measures convergence along two fundamental dimensions: depth and breadth. For depth calculation, we use IPC textual descriptions as the analytical foundation and enhance this assessment by incorporating supplementary patent metadata into a heterogeneous graph structure. This graph is then modelled using Heterogeneous Graph Transformers (HGT) in combination with Sentence-BERT (SBERT), enabling a precise representation of knowledge integration across technological boundaries. Complementing this, the breadth dimension captures the diversity of technological fields involved, quantified through the Shannon Diversity Index (SDI) to measure the variety of technological combinations within patents. Our final TCI is constructed using the Entropy Weight Method (EWM), which objectively assigns weights to both dimensions based on their information entropy. To validate our approach, we compare the proposed TCI against established convergence measures, demonstrating its comparative advantages. We further establish empirical reliability through a novel robustness test that regresses TCI against indicators of patent quality. Applying this framework to Chinese patents related to the twin transition (2003-2024) reveals that technological convergence has a significant positive effect on patent quality, confirming that higher levels of technological convergence are associated with higher-quality innovations. These findings are further substantiated through comprehensive robustness checks. Our multidimensional approach provides valuable practical insights for innovation policy and industry strategies in managing emerging cross-domain technologies.

---

[*] Corresponding Author.
*Email address*: *julielcj@163.com*

# 1. Introduction

*Technological convergence* describes the phenomenon where boundaries between technological areas and disciplines are increasingly blurred, enabling the integration of previously distinct domains (Curran & Leker, 2011; Rosenberg, 1963). This increasingly significant concept has drawn widespread attention from policymakers, practitioners, and researchers, given its far-reaching implications across multiple sectors and its role as a catalyst for innovation and industrial evolution (Gauch & Blind, 2015; Jeong et al., 2015). Researchers have identified a rising prevalence of convergence in recent decades through analyses of patent data showing growing overlap among previously separate technology sections, indicating that formerly distinct technologies are increasingly co-invented or used together (Lee et al., 2023). Furthermore, technological convergence creates significant opportunities for breakthrough innovations that emerge specifically at the intersection of different fields, potentially generating entirely new products and industries (Huang et al., 2020; Zhang et al., 2025).

However, the fundamentally multidimensional nature of technological convergence still raises methodological challenges for precise measurement. For example, technological convergence can deepen (increase knowledge integration intensity within closely related trajectories) and broaden (expand across distinct technological fields), ultimately producing richer cross-domain technologies (Luan et al., 2021). Early studies built a valuable foundation by exploiting statistical relationship-based methods such as IPC co-occurrence statistics (Tang et al., 2020; Yun & Geum, 2019), co-word analysis (Lee et al., 2015), and traditional diversity indices like the Shannon Diversity Index (SDI) and the Herfindahl-Hirschman Index (HHI) (Lu et al., 2017; Zhu et al., 2022). These techniques effectively depict breadth while capturing depth only through basic co-occurrence patterns that miss the semantic intensity of knowledge integration across fields, thus lacking the resolution needed to uncover fine-grained cross-domain relationships (Borés et al., 2003; Kim et al., 2014).

More recently, AI-driven approaches leveraging semantic embedding such as Word2Vec (Hong et al., 2022) and BERT (Giordano et al., 2021), along with graph-based models like Graph Convolutional Networks (GCN) (Zhu & Motohashi, 2022) and Heterogeneous Graph Transformers (HGT) (Jiang et al., 2024) offer enhanced capabilities. These advanced methods identify semantic relationships and structural interconnections within large-scale patent datasets more effectively than earlier statistical methods (Gozuacik et al., 2023; Yang et al., 2024). They also facilitate temporal analyses by modeling how knowledge flows evolve over time (Yun & Geum, 2019), as shown by studies using dynamic network analysis to trace the longitudinal evolution of technological relationships (Choi et al., 2018; Kim et al., 2014). However, while these studies have advanced our understanding of technological convergence, many have approached depth and breadth dimensions separately, suggesting valuable research opportunities that explore how these complementary aspects might work in concert to shape technological convergence patterns. A comprehensive assessment framework incorporating both depth and breadth dimensions offers significant advantages for technological convergence analysis. Approaches that focus primarily on breadth may tend to emphasize technological diversity without fully accounting for the quality of connections between fields (Papazoglou & Spanos, 2018). Similarly, methodologies centered exclusively on depth might not fully recognize valuable innovations that span traditional domain boundaries (Park & Yoon, 2018). By integrating these complementary perspectives, researchers can develop a more balanced and operationally relevant understanding of technological convergence patterns.

To address this gap, we construct an AI-enhanced Technological Convergence Index (TCI), combining semantic analysis and heterogeneous graph learning to measure technological convergence across both depth and breadth dimensions. Our approach incorporates two learned modules that work collaboratively: (1) a transformer-based sentence encoder (Sentence-BERT, SBERT) that maps IPC descriptions, patent titles, and abstracts into contextual embedding vectors; and (2) a Heterogeneous Graph Transformer (HGT) that performs relation-aware attention over a patent–IPC–title/abstract–applicant graph to produce structure-aware node representations.

We compute depth from these learned representations through a two-step process. The core divergence between the primary IPC and each secondary IPC is measured as cosine dissimilarity in the fused embedding space, where SBERT text embedding are refined through HGT message passing. The peripheral heterogeneity among secondary IPCs is then quantified as the attention-weighted average pairwise dissimilarity, with weights derived from HGT's relation-specific attention mechanisms. This design enables depth to reflect both semantic proximity through SBERT and topological context through HGT, yielding a comprehensive measure of cross-domain knowledge integration. Compared with traditional convergence measures based on IPC co-occurrence counts or text-only similarity, this AI-enhanced depth jointly learns semantics and structure through relation-aware HGT, capturing cross-domain integration signals that single-view methods typically miss.

Breadth is quantified using the Shannon Diversity Index (SDI) to capture the variety of IPC portfolios within each patent. This component does not involve AI modules but provides an established measure of technological diversity. The Entropy Weight Method (EWM) aggregates depth and breadth by assigning data-driven weights based on their information entropy. EWM serves as an objective weighting mechanism rather than an AI component, ensuring that the final index reflects the relative information content of each dimension.

Seven representative baselines are selected for validation: **two** IPC co-occurrence indicators that infer proximity solely from adjacency statistics; **one** transformer-based semantic model that identifies latent textual similarity while not accounting for network structure; **one** topology-oriented graph model that learns heterogeneous connections without incorporating an explicit diversity signal; and **two** hybrid variants that combine a single-dimension depth (or semantic) score with a Shannon-based breadth metric, though still treating these dimensions as separate components; and **one** extension using a Rao-Stirling diversity index that incorporates semantic embedding distances into breadth measurement, avoiding potential overestimation when categories are semantically close.

In addition, we propose a novel robustness test that serves as an external benchmark, examining the relationship between technological convergence and patent quality, thereby validating the empirical reliability and practical applicability of our approach. This test aligns with the knowledge-recombination perspective, which suggests that stronger technological convergence contributes to enhanced technology quality (Zhao et al., 2023). Moving beyond conventional baseline comparisons, we establish this relationship by regressing TCI against patent quality metrics, providing additional verification of our measurement framework's effectiveness.

We apply the TCI to Chinese patents associated with the twin transition technologies (2003-2024). The twin transition technologies, initially advanced by the European Union as a route to a carbon-neutral economy by 2050 (European Commission, 2022), highlight how digital technologies can accelerate green transformation (Ortega-Gras et al., 2021). China's rapidly expanding portfolio in this area offers an ideal test bed (Brueck et al., 2025). It is sizable enough for robust statistical analysis, inherently interdisciplinary, and strategically aligned with global sustainability goals. By revealing how depth and breadth converge within these patents, our study elucidates the structural mechanics underlying China's green-digital trajectory and provides globally transferable evidence for policymakers, industry leaders, and international bodies seeking to leverage digital innovation for net-zero objectives (Fouquet & Hippe, 2022; Myshko et al., 2024).

The remainder of this paper is structured as follows. Section 2 reviews prior studies on technological convergence and its measurement. Section 3 outlines the proposed research framework and methodology. Section 4 presents a case study to evaluate the stability and validity of the proposed TCI. Finally, Section 5 concludes the paper with a discussion of the developed index and potential directions for future research.

## 2. Literature review

### 2.1 Technological convergence

Technological convergence refers to the process of integrating previously distinct technological domains, scientific knowledge, and markets to create new solutions and innovations (Borés et al., 2003; Caviggioli, 2016). This process involves technology selection, combination, and integration, leading to the blurring of boundaries between different fields (Guo et al., 2022; Luo & Zor, 2022). The concept of technological convergence was first introduced by Rosenberg (1963) in his study on the *Technological Changes in the Machine Tool Industry (1840-1910)*. He used the term to contrast converging technological trajectories with sequences of parallel and independent activities (Rosenberg, 1963). Since then, scholars from various disciplines have expanded upon this foundational concept, providing a substantial body of theoretical and empirical evidence that enhances our understanding of technological convergence (Hussain et al., 2022; Lee et al., 2023).

Two main perspectives have emerged as particularly influential in the technological convergence literature. Boundary spanning represents one significant viewpoint, with scholars such as Curran and Leker (2011) and Kim et al. (2015) describing it as the facilitation of knowledge exchange across disciplinary, organizational, or technological domains, which promotes innovation through external engagement. Collaborative efforts across boundaries enable participants to access novel perspectives and complementary capabilities, thereby overcoming the limitations of established knowledge or institutional structures (Hsiao et al., 2012; Kark et al., 2015).

Knowledge recombination constitutes the second major perspective. Researchers such as Fleming and Sorenson (2001) and Singh and Fleming (2010) propose that innovation frequently emerges from the novel reconfiguration of existing knowledge elements. This perspective illustrates how combining seemingly unrelated or previously isolated knowledge fragments can produce breakthrough outcomes, particularly under conditions of high uncertainty (Gruber et al., 2013; Zhong et al., 2024). In this view, innovation involves not merely accessing diverse knowledge, but also creatively restructuring that knowledge to address emerging needs (Savino et al., 2017; Xiao et al., 2022).

Technological fusion, a related but distinct notion, complements convergence by emphasizing how domains are combined at the component and architectural levels. Whereas convergence describes the portfolio-level coalescence of distinct knowledge fields, fusion concerns the interoperability and functional integration of elements within or across systems. Empirically, fusion is often operationalized via IPC co-classification and structure-aware network features that reveal cross-domain linkages and architectural blending; it has also been linked to firms' strategic repositioning as industries evolve under convergence. However, because fusion focuses more strongly on component-level integration while our study emphasizes portfolio-level diversity and cross-domain variety, we adopt convergence as the central lens for developing our multidimensional TCI.

Building on these perspectives, this study integrates boundary spanning and knowledge recombination to explain how cross-field knowledge flows and recombinative search jointly drive convergence across technological domains. Boundary spanning facilitates the flow of knowledge across different technological fields, enabling the integration of diverse knowledge sources, while knowledge recombination leverages these heterogeneous knowledge assets to generate new technological possibilities (Rosenkopf & Nerkar, 2001). Boundary spanning and knowledge recombination play distinct yet complementary roles in the process of technological convergence.

## 2.2 Measurement of technological convergence

One of the primary challenges in studying technological convergence lies in establishing reliable measurement standards (Gauch & Blind, 2015). Effective measurement provides a foundational basis for technological innovation, industrial advancement, policy formulation, and patent management (Choi et al., 2015; Lei, 2000). Yet, developing measurement methods that both reflect the essence of technological convergence and support rigorous quantitative analysis remains complex (Thorleuchter et al., 2010). Consequently, researchers have explored a range of strategies to capture the multifaceted nature of convergence while ensuring the feasibility of empirical investigation.

The first and most fundamental step is the organization and selection of data to develop effective measurement approaches for technological convergence. Patent data, with its systematic, comprehensive, and hierarchical nature, has become an ideal data source for studying technological convergence (Caviggioli, 2016; Kim et al., 2017). Scholars tend to use the International Patent Classification (IPC) for technical classification in technological convergence studies (Leydesdorff et al., 2014), as its technical orientation better captures the essence of patents compared to industry-based classifications (Harris et al., 2010). Thus, patent data and IPC classification have become foundational for technological convergence measurement.

Scholars have proposed a variety of IPC-based measurement approaches that can be broadly divided into two categories. One focuses on technological field diversity, using indicators like the HHI (Lee, 2023; Lu et al., 2017), SDI (Jung et al., 2021; Zhu et al., 2022), and Rao-Stirling Diversity (Leydesdorff et al., 2019) to quantify the breadth and dispersion of field combinations. The other examines similarities among technological fields through measures such as Jaccard similarity (Giordano et al., 2021), cosine similarity (San Kim & Sohn, 2020), and graph-based similarity, uncovering overlaps that may drive convergence. While these IPC-based methods provide valuable cross-sectional insights, many offer static snapshots rather than capturing the temporal evolution of convergence.

Temporal and structural dynamics received greater attention with the introduction of co-occurrence-based methods such as patent co-classification statistics (Choi et al., 2015) and co-word analysis (Seo et al., 2012). As these approaches still tended to focus on specific time points, researchers developed more sophisticated methods incorporating longitudinal data. Dynamic methodologies emerged, including dynamic network analysis (Giordano et al., 2021), panel data analysis (Utku-İsmihan, 2019), and social network analysis (Han & Sohn, 2016; Luan et al., 2013), which track shifts in technological convergence over time. Alongside these developments, knowledge-flow approaches such as citation network analysis gained prominence (Kim et al., 2014; Zhang et al., 2017), illuminating how patents or papers influence each other across different domains and illustrating the evolving pathways of technological integration (Mejia et al., 2021). Beyond descriptive links, causal tools such as Granger causality tests and Difference-in-Differences (DID) designs are increasingly applied to identify the directional impact of convergence on economic or innovation outcomes (Guo & Zhong, 2022; Luan et al., 2022).

Recent advances in semantic analysis techniques have further refined technological convergence measurement. Word embedding models such as Word2Vec and GloVe generate vector representations that capture semantic relationships between words, enabling more nuanced quantitative analyses of similarities in patents or publications (Hong et al., 2022; Lee et al., 2022; Zhang et al., 2018). Transformer-based architectures like BERT provide a deeper contextual understanding of technical terminology and its interconnections across diverse fields (Song et al., 2023; Wang et al., 2023; Zhu & Motohashi, 2022). Additionally, topic modeling approaches, particularly Latent Dirichlet Allocation (LDA) (Cho et al., 2021; Song & Suh, 2019), help identify latent themes in large document collections, revealing how previously distinct technological domains converge or diverge over time. Where multiple indicators coexist, composite or entropy-weighted indices have been proposed to synthesize breadth, similarity, and network dimensions into a single convergence score (Lee et al., 2021). **Table. 1** illustrates the comparative framework of technological convergence measurement approaches.

**Table. 1 Comparative Framework of Technological Convergence Measurement Approaches**

| Method | Dimensional Scope | Data Dependency | Main Strengths | Limitations / Biases |
|---|---|---|---|---|
| **Diversity Indicators** | Single-dimensional (breadth) | IPC shares; Rao-Stirling requires embedding distances | Simple, interpretable; captures portfolio variety; RS incorporates semantic distance | SDI ignores semantic proximity (may overestimate convergence); HHI overemphasizes concentration |
| **Co-occurrence / Co-classification** | Mostly single-dimensional | IPC co-occurrence matrix, keywords | Straightforward; intuitive measure of adjacency | Miss latent semantics; sparse for emerging tech; sensitive to classification practice |
| **Citation / Knowledge Flow** | Single or multi-dimensional | Citation networks, bibliographic coupling, co-citation | Directional, traces knowledge transfer | Lagging indicator; influenced by citation behaviour and norms |
| **Semantic Embedding** | Single-dimensional (depth) or input to composites | Full text, titles, abstracts | Captures latent semantic proximity; robust to language | Ignores network structure; domain bias in corpora |
| **Network / Topological Metrics** | Single or multi-dimensional | Heterogeneous links (patent–IPC, IPC–IPC, etc.) | Reveals structural position and cross-domain bridges | Lacks semantic content; results sensitive to network design |
| **Dynamic Models** | Multi-dimensional | Time-stamped co-occurrence, citation, or network data | Captures path dependence and life-cycle effects | Requires long, consistent data; complex to estimate |
| **Composite / Weighted Combinations** | Multi-dimensional | Combination of semantic, network, and diversity features | Integrates complementary dimensions; data-driven weights (e.g., entropy) | Interpretability depends on clarity of components; sensitive to weighting scheme |

## 2.3 Research gaps and objectives

Although measurement techniques for technological convergence have progressed significantly, three methodological challenges persist in current research approaches. First, existing studies typically analyze breadth and depth as independent constructs. Diversity-oriented indicators assess breadth, while similarity-oriented methods evaluate depth. Few frameworks incorporate these dimensions within a unified composite index, leaving the relationship between knowledge-portfolio diversity and cross-domain integration intensity largely unexamined. Second, recent innovations in natural language processing and graph learning, such as transformer embedding and heterogeneous graph networks, are predominantly applied to similarity estimation. When diversity information is included, it often appears as an ex-post descriptor, maintaining analytical separation between these dimensions. Third, many studies lack robustness testing that connects convergence measures to practical application metrics. Without such validation, an index might identify statistical patterns but fail to demonstrate practical significance, potentially leading to questionable conclusions and diminishing its value for innovation management and policy development.

To address these limitations, we develop a TCI that evaluates depth and breadth simultaneously within a unified framework. Our approach leverages recent advances in AI-enhanced methods by integrating two learned modules. First, SBERT embedding capture fine-grained textual similarity across IPC descriptions, titles, and abstracts. Second, a HGT models relation-aware structures linking patents, IPCs, and applicants to produce structure-aware node representations. This dual-module architecture enables us to jointly account for semantic proximity and structural integration, representing an advance over prior single-view measures.

We derive depth measurements from this HGT-SBERT heterogeneous patent graph by quantifying the semantic strength of boundary-spanning connections. Breadth is assessed using the SDI to reflect the variety of IPC-based knowledge combinations. The EWM objectively weights these dimensions based on their information entropy, and we verify the index's practical relevance through regression analysis against patent quality indicators. This multidimensional, validated framework offers scholars and policymakers a comprehensive method for assessing technological convergence across innovation systems.

## 3. Methodology

### 3.1 Research framework

Technological convergence fundamentally operates as a core-periphery recombination process. In this process, inventions simultaneously extend the core knowledge represented in their main classification while incorporating peripheral knowledge from auxiliary fields. Based on established literature examining knowledge coherence and the concepts of related versus unrelated variety, we enhance our approach by disaggregating "depth" into two complementary facets before integrating it with "breadth". This refined design enables us to measure both the extent to which an invention diverges from its core domain and the heterogeneity of that divergence, establishing a theoretically sound foundation for our TCI.

To clarify how these dimensions map onto established innovation theory, we situate our framework within the literatures on knowledge relatedness, cognitive distance, and path dependence. In this tradition, innovation outcomes hinge on balancing the exploitation of related knowledge (coherence and depth) with the exploration of distant knowledge (breadth and variety). Accordingly, Depth represents the intensity and coherence of cross-domain integration, closely aligned with relatedness and coherence, whereas Breadth captures the diversity of a firm's technological portfolio, shaping the opportunity set for recombination. This grounding provides a robust theoretical rationale for assessing convergence along both dimensions simultaneously.

Our analysis therefore examines technological convergence through two primary dimensions. When an invention demonstrates a higher TCI value, this indicates both greater cross-field scope (depth) and increased technological diversity (breadth). **Fig. 1** illustrates the comprehensive theoretical framework that guides this study, showing how these dimensions interact to provide a holistic measurement of technological convergence patterns.

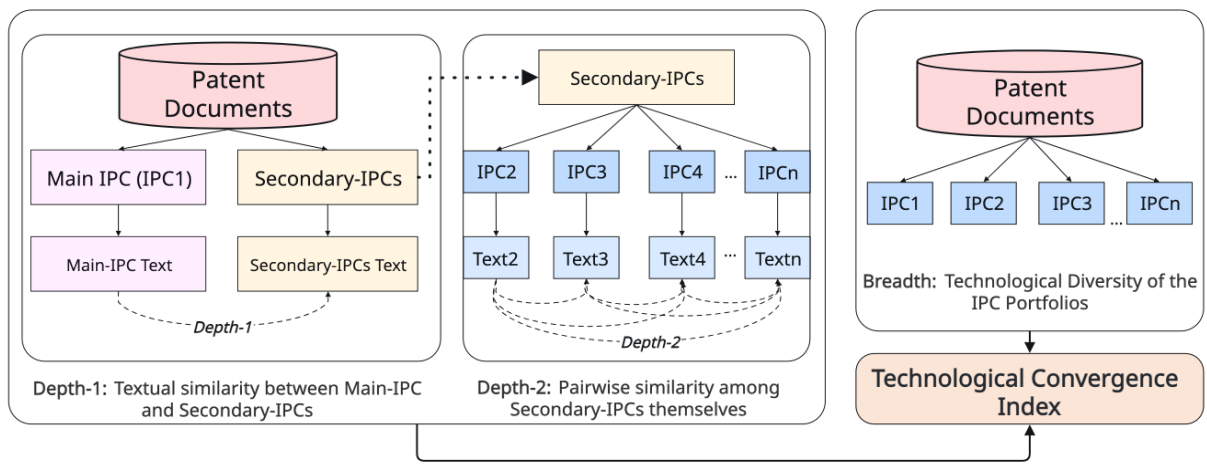

**Fig. 1 Theoretical framework**

As shown in **Fig. 1**, to conduct a more detailed study on *Depth*, our approach is divided into two steps. **Step 1** focuses on the similarity between a patent's main-IPC and its secondary-IPCs (*Depth-1* in **Fig. 1**). In patent analysis, the main-IPC refers to the primary classification code of a patent, indicating the main technological field the patent belongs to. IPC system classifies patents into multiple codes, with the main-IPC designating the core technological direction of the patent, typically assigned by the patent examiner or applicant. Computing the similarity between the main-IPC and secondary-IPCs helps assess the extent of cross-field technological extension around the core technology field of patent. If the similarity is consistently high, it suggests that the patent remains concentrated within its core field. Conversely, a lower similarity suggests that the patent extends beyond its core domain, leading to a greater depth of technological convergence.

**Step 2** examines the similarity among secondary-IPCs (*Depth-2* in **Fig. 1**) as a supplementary analysis to **Step 1**. If the secondary-IPCs are highly similar to each other, then the patent's auxiliary technological fields form a closely related technology cluster. Conversely, a low similarity among secondary-IPCs indicates that the patent encompasses a broader set of cross-field technologies. Through these two steps, one can determine both the extent of cross-field expansion relative to the core field (*Depth-1*) and how wide-ranging the secondary IPCs are (*Depth-2*). We then integrate *Depth-1* and *Depth-2* to derive an overall *Depth* measure. Lower IPC similarity values at each step yield higher cross-field *Depth*, reflecting more pronounced multi-field convergence.

*Breadth* measures the diversity in patent technology distribution. The calculation of technological convergence breadth assesses how various IPC classifications are distributed within a patent, reflecting the range of technological fields it encompasses. Patents typically contain multiple IPC codes, which collectively represent the patent's technological diversity. Patents with numerous, evenly distributed IPC categories demonstrate higher technological convergence breadth, indicating integration across diverse technological fields. Conversely, patents with few, concentrated IPC categories exhibit lower technological convergence breadth, suggesting a focus on limited technological domains.

Finally, we employ the EWM to weight and integrate *Depth* and *Breadth*, constructing our TCI.

## 3.2 Measuring the depth of technological convergence

The measurement of Depth is based on IPC similarity, which integrates both semantic and graph structural information. Specifically, we construct a heterogeneous graph that includes patents, IPCs, applicants, and topics as interconnected entities. We then leverage a HGT framework in conjunction with SBERT to perform comprehensive representation learning on this graph. Once node embedding are learned, we calculate inter-IPC similarities and transform those similarity values to represent cross-field depth. **Fig.2** illustrates the framework for measuring the depth of technological convergence.

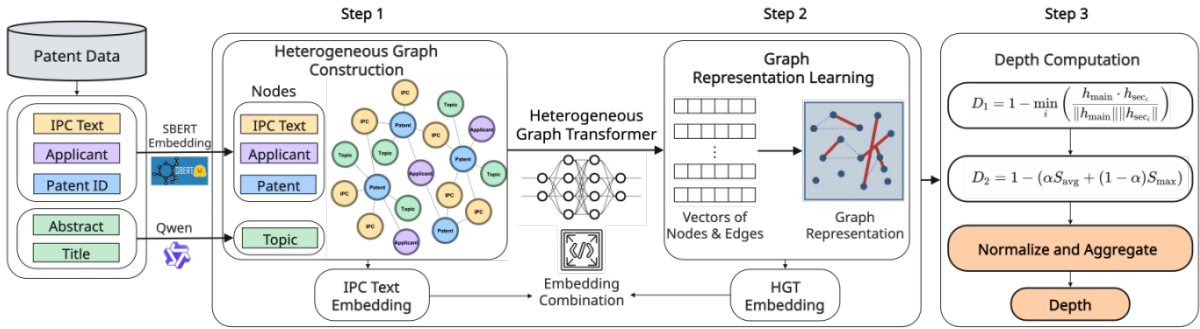

**Fig. 2 Framework for computing the depth of TCI**

*Notes:*
*1. Patent ID refers to the patent application number, used as the unique identifier for each patent in the graph construction.*
*2. IPC text represents the descriptive text of each patent's International Patent Classification (IPC) at the subgroup level. For instance: A01B SOIL WORKING IN AGRICULTURE OR FORESTRY; PARTS, DETAILS, OR ACCESSORIES OF AGRICULTURAL MACHINES OR IMPLEMENTS, IN GENERAL, including references to related subgroups (e.g., A01C 5/00; A01D 42/04). Official IPC subgroup definitions are adopted.*

### *3.2.1 Heterogeneous graph construction*

Our heterogeneous graph incorporates four distinct node types: *Patent, IPC, Topic,* and *Applicant*. Patent nodes function as central elements within the graph structure, linking to other node types while retaining crucial attributes. IPC nodes represent specific technological classification codes, further distinguished into main and secondary IPCs to display each patent's primary and auxiliary technical domains. *Applicant* nodes are extracted from standardized patent documents to identify inventors responsible for patent applications. Topic nodes represent the technical topics of each patent, extracted by applying the Qwen 2.5-7B (QwenTeam, 2019) large language model to patent titles and abstracts to obtain 10 representative topics. Qwen is selected for its ability to be deployed entirely offline, strong bilingual coverage of Chinese and English technical terminology, capacity to handle long patent texts, and efficient operation on consumer-grade GPUs (Xu et al., 2024), essential advantages for large-scale, privacy-sensitive patent analytics that provide richer semantic insights than traditional NER-based approaches (Bai et al., 2023).

We employ SBERT to generate initial feature vectors for text-based nodes (e.g. *IPC*, *Applicant*, and *Topic*). SBERT is specifically fine-tuned for sentence-level similarity tasks, producing high-quality cross-lingual semantic vectors and delivering orders-of-magnitude faster inference than models that require additional fine-tuning for similarity estimation, which is critical when calculating millions of IPC pairwise similarities (Sun et al., 2022). For each text-based node $i$:

$$x_i^{(0)} = SBERT(Text_i) \tag{1}$$

where $Text_i$ denotes the node's textual attribute. These initial feature vectors serve as the foundation for subsequent graph construction.

Edges in our graph represent meaningful relationships: *Patent–IPC* edges define the technical scope, *IPC–IPC* edges quantify technical similarity among different IPC categories, *Patent–Topic* connect patents to their extracted semantic topics, creating a concise representation of thematic content, and *Applicant–Patent* edges represent the link between innovators and their intellectual property, enabling the analysis of organizational contributions to technological advancement. Formally,

$$G = (V, E, R) \tag{2}$$

where $V$ is the set of nodes, $E$ the set of edges, and $R$ the set of relation types. With the graph structure established, we embed node representations that capture both semantic and structural properties.

*3.2.2 Graph representation learning*

To effectively learn node embedding from our heterogeneous graph structure, we utilize HGT as the foundational architecture. HGT is specifically designed to address the complexity of multi-type graphs, each containing distinct relationships and connectivity patterns. The pre-computed SBERT embedding serve as the initial node features, which are then enriched with structural information through the HGT embedding process. This approach preserves the original textual semantics while incorporating domain-specific relationships from the graph structure.

During training, HGT updates node embedding through message passing, where each node aggregates information from neighbors with relation-specific weighting. The node embedding at layer $l+1$ for node $i$ is defined as:

$$h_i^{(l+1)} = \sum_{r \in R} \sum_{j \in N_r(i)} \alpha_{ij}^r W^r h_j^{(l)} \tag{3}$$

Where $N_r(i)$ represents node $i$'s neighbors connected via relation $r$, $W^r$ is the relation-specific transformation matrix, and $\alpha_{ij}^r$ is the attention coefficient quantifying neighbor $j$'s importance. This process integrates local and global graph contexts, producing embedding that encode both the initial textual semantics and the graph's structural relationships.

After training, each node $i$ output a final embedding $h_i$. For IPC nodes, we refine representation quality by combining the HGT-derived structural embedding $\tilde{h}_{IPC}$ with the original SBERT embedding $x_i^{(0)}$:

$$h_{IPC} = concat(\tilde{h}_{IPC}, x_{IPC}^{(0)}) \tag{4}$$

Where $\tilde{h}_{IPC}$ captures network-structured information, and $x_{IPC}^{(0)}$ retains the original semantic features, creating a representation that balances both aspects for subsequent similarity calculations.

### 3.2.3 Depth Calculation

Using the fused IPC embedding, we quantify technological convergence depth through inter-IPC similarity calculations, integrating both semantic representations and graph structural information. *Depth-1* using the minimum cosine similarity between the main-IPC and each of its secondary-IPC.

$$S_{main,sec} = min_i \left( \frac{h_{main} \cdot h_{sec_i}}{\|h_{main}\| \|h_{sec_i}\|} \right) \tag{5}$$

*Depth-2* captures the extent of cross-field integration among the secondary-IPCs. It is calculated using a dynamically weighted similarity measure.

$$S_{sec,sec} = \alpha S_{avg} + (1-\alpha) S_{max} \tag{6}$$

$$S_{avg} = \frac{1}{M} \sum_{(i,j) \in P} \left( \frac{h_{sec_i} \cdot h_{sec_j}}{\|h_{sec_i}\| \|h_{sec_j}\|} \right) \tag{7}$$

where $h_{main}$ represents the embedding vector of the main-IPC, $h_{sec_i}$ represents the embedding vector of the $i_{th}$ secondary-IPC. The minimum value ($min_i$) ensures that *Depth-1* captures the maximum cross-field expansion, i.e., the secondary-IPC that is least similar to the core domain. In *Depth-2*, $P$ is the set of all possible secondary IPC pairs and $M$ is the number of such pairs. $S_{avg}$ is the mean pairwise cosine similarity between all secondary-IPCs, while $S_{max}$ is the maximum pairwise similarity among secondary-IPCs. The dynamic weight $\alpha = \frac{n}{n+k}$ is determined by the number of secondary IPCs, with smoothing parameter $k$.

To obtain the final *Depth* scores, we transfer the similarity measures as follows:

$$D_1 = 1 - S_{main,sec} \tag{8}$$

$$D_2 = 1 - S_{sec,sec} \tag{9}$$

$$Depth = \omega_1 \cdot D_1 + \omega_2 \cdot D_2 \tag{10}$$

where $D_1$ and $D_2$ are the scores of *Depth-1* and *Depth-2*, respectively. The final *Depth* score is a weighted sum of the two components, with $\omega_1$ and $\omega_2$ representing entropy-based weights that ensure a balanced contribution from both terms.

### 3.3 Measuring the breadth of technological convergence

*Breadth* reflects the diversity of technological fields covered by a patent, indicating whether it spans a broader range of domains based on its IPC classifications. To quantify the diversity of IPC categories, we define *Breadth* based on the normalized SDI, calculated as:

$$Breadth = SDI_{norm} = \frac{SDI - SDI_{min}}{SDI_{max} - SDI_{min}} \tag{11}$$

where $SDI = -\sum_{i=1}^{N} p_i \ln p_i$, with $p_i = \frac{n_i}{\sum_{j=1}^{N} n_j}$ representing the proportion of IPC categories $i$, and $N$ is the total number of IPC categories. $SDI_{max}$, $SDI_{min}$ refer to the maximum and minimum SDI values observed across all patents.

### 3.4 Construction of technological convergence index

The construction of the TCI begins with determining the weights of *Depth-1* ($D_1$), Depth-2 ($D_2$), and *Breadth* ($D_3$). These weights are calculated using the EWM, which objectively reflects the relative importance of each indicator based on its variability across the dataset. We first normalize the values to ensure comparability. Then, we compute the entropy $E_j$ for each indicator $j$ as follows:

$$E_j = -\frac{1}{\ln N} \sum_{i=1}^{N} p_{ij} \ln p_{ij} \tag{12}$$

where $p_{ij}$ represents the proportion of the $i_{th}$ patent's value for indicator $j$ relative to all patents, and $N$ is the total number of patents. Lower entropy values indicate higher variability in the data and thus a greater weight for that indicator. The final weight for each component is calculated as:

$$\omega_j = \frac{1 - E_j}{\sum_j (1 - E_j)} \tag{13}$$

We impose a constraint that the weight of $D_1$ should always be greater than $D_2$ ($\omega_1 > \omega_2$), reflecting the idea that cross-field expansion relative to the main IPC is a stronger indicator of convergence than intra-field variation among secondary IPCs. If $\omega_2$ exceeds $\omega_1$, an adjustment is applied, followed by re-normalization to ensure that all weights sum to 1. The final TCI is calculated as:

$$TCI = \omega_1 \cdot D_1 + \omega_2 \cdot D_2 + \omega_3 \cdot D_3 \tag{14}$$

A higher TCI value indicates that a patent spans multiple technological fields, integrates a diverse range of technologies, and exhibits strong technological convergence. Conversely, a lower TCI value suggests that a patent remains concentrated within its core field, with limited cross-field expansion and lower technological diversity.

## 4. Case study: Empirical study of TCI in the twin transition

Our case study uses China's twin transition technology patent data from 2003 to 2024 as the empirical analysis sample. We provide a descriptive analysis of the TCI in the context of twin transition patents while empirically validating its effectiveness by examining its relationship with patent quality. This integrated approach provides deeper insights into how TCI captures cross-domain interactions. As a result, it guides researchers and practitioners in leveraging digital and green synergies to drive innovation.

### 4.1 Background and data

*Twin transition* represents an integrated process where digital and green transformations co-occur within technological and industrial fields (Fouquet & Hippe, 2022). Previous studies suggest that digitalization and greening are interconnected, mutually reinforcing, and co-evolving processes (Tabares et al., 2025). Digital technologies provide powerful tools for monitoring, managing, and optimizing green objectives (Mondejar et al., 2021; Wu et al., 2021). Sustainability principles guide digital innovations toward lower-carbon and environmentally friendly directions (Bhatia et al., 2024). This interplay enhances the sustainable competitiveness of organizations. Many global economies like the European Union have increasingly adopted the twin transition, aiming to leverage digital technologies to support green transformation and establish more efficient, resilient, and sustainable economic models (Garito et al., 2023; Salvi et al., 2022).

Twin transition demonstrates typical features of technological convergence, as it spans multiple disciplines such as economics, environmental sciences, engineering, and policy studies (Paiho et al., 2023). These cross-disciplinary characteristics align with our research objectives and allow us to identify technological convergence patterns with practical relevance at the same time. Patents serve as critical indicators of technological innovation, systematically reflecting innovative activities and developmental trends within specific technological fields (Caviggioli, 2016; Grupp, 1994). Recently, there has been a surge in patent data related to the twin transition, providing us with comprehensive data to conduct this study. Therefore, we adopt patent data to empirically investigate the interactive relationship between digitalization and greening from the technological innovation perspective.

Specifically, we apply the Cooperative Patent Classification (CPC) system to identify and capture relevant green technology patents, focusing on the CPC subfields Y02A, Y02B, Y02C, Y02D, Y02E, Y02P, Y02T, Y02W, and Y04S at the initial stage (EPO, 2022). Subsequently, we filtered digital technology patents based on *China's Digital Economy Core Industries and International Patent Classification Comparison Table* (CNIPA, 2023) to construct our original dataset. After comprehensive data collection, classification, and cleaning procedures, we obtained a final dataset comprising 87,795 twin transition-related patents filed in China from 2003 to 2024.

## 4.2 Descriptive Analysis of TCI

### 4.2.1 TCI distribution and overall structure

**Fig. 3** and **Fig. 4** illustrate the comprehensive distribution and structural characteristics of the TCI within twin transition patents. **Fig. 3** presents the density distribution of the TCI from 2003 to 2024, illustrating the evolution of technological convergence over the past 22 years. The visualization reveals a distinct temporal shift. In the earlier period (2003-2010), depicted in blue tones, the distribution leans toward higher TCI values, typically ranging from 0.3 to 1.0, indicating a phase of more balanced and widespread technological integration. In contrast, the later years (2011-2024), shown in red and orange, exhibit a clear bimodal distribution, with a sharp concentration near zero and a secondary peak between 0.15 and 0.20. This transformation suggests a growing polarization in convergence patterns: recent patents increasingly reflect either highly specialized technologies (with minimal convergence) or moderately integrated innovations. Earlier patents, by comparison, are more uniformly distributed across higher TCI values, pointing to a historically more consistent convergence profile.

**Fig. 4** illustrates the density distribution of the TCI across major IPC sections, revealing substantial variation in convergence patterns among technological domains. Sections G (Physics) and H (Electricity), highlighted in red and orange, exhibit multimodal distributions with a sharp peak near zero, reflecting a high concentration of highly specialized patents. Additional peaks are observed around 0.15 and in the range of 0.2 to 0.25, with the overall spread extending up to 0.4. In contrast, Sections A (Human Necessities), B (Performing Operations; Transporting), and C (Chemistry; Metallurgy), shown in shades of blue, display unimodal distributions centred around 0.3 to 0.4, indicating a more balanced and stable pattern of convergence across knowledge domains. Notably, high TCI values (above 0.6) are rare across all sections, suggesting that extreme convergence is uncommon. These distributional patterns point to structural differences in technological convergence across fields: while Sections G and H reflect both specialization and localized integration, Sections A, B, and C demonstrate more consistent, multi-domain integration.

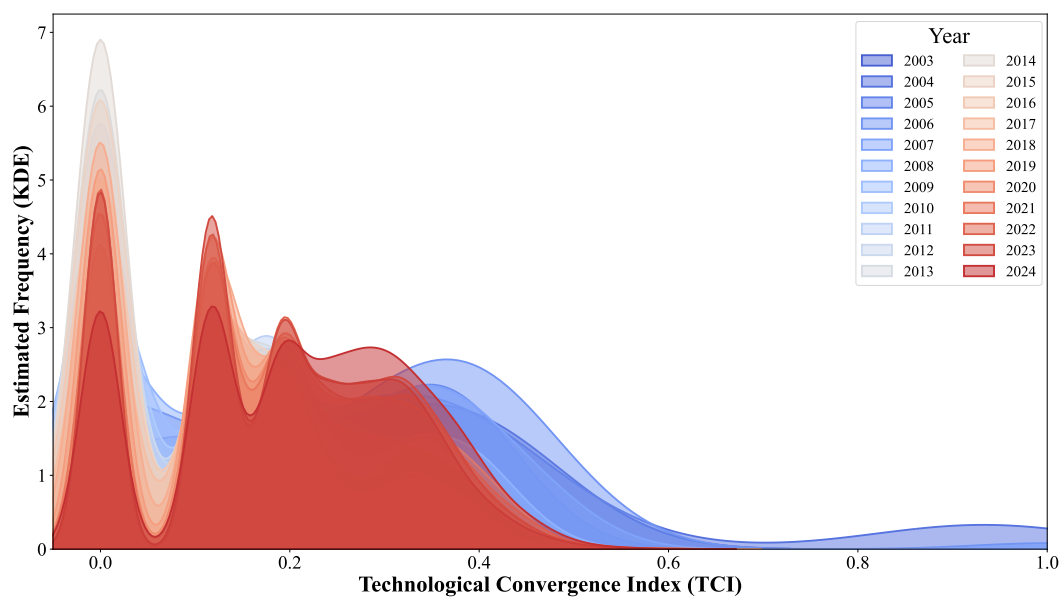

*Note: Each curve shows the estimated distribution of TCI for a specific year using kernel density estimation (KDE).*
**Fig.3 Year-wise density distribution of TCI (2003–2024)**

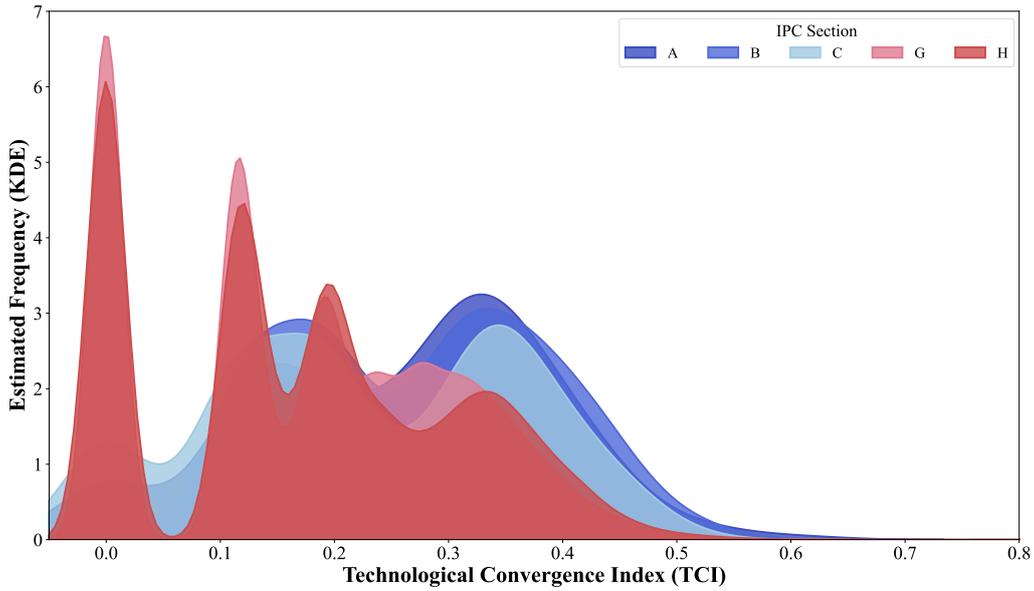

*Note: Each curve shows the estimated distribution of TCI for a given IPC section using kernel density estimation (KDE).*

**Fig.4 IPC-sectional density distribution of TCI**

### *4.2.2 Development trends of TCI*

**Fig.5a** and **Fig.5b** focus on the temporal and sectoral developmental trends of TCI. Investigating these dynamics clarifies how technological convergence evolves over time and across different technology areas. **Fig. 5a** highlights TCI's developmental trajectory, emphasizing a noticeable upward trend from 2003 to 2024. Especially after around 2010, a rapid increase in TCI implies accelerating integration, likely reflecting favorable innovation policies and market incentives promoting digital and green synergies. **Fig. 5b** further dissects sectoral differences, showcasing varied IPC-sectional TCI growth trajectories. Some sectors exhibit rapid convergence, while others grow more gradually, indicating uneven responsiveness to twin transition incentives and policies.

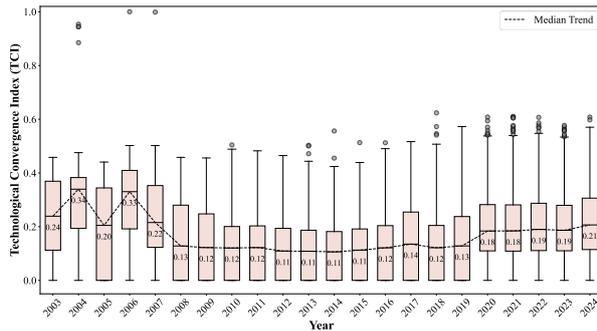 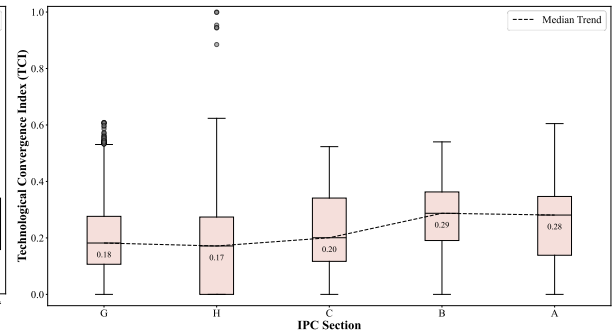

Fig.5a Year-wise development trends in TCI (2003-2024)    Fig.5b IPC-sectional development trends of TCI

**Fig.5 Development trends of TCI**

### **4.3 Results of the application of TCI**

Building on the construction and theoretical rationale of the TCI presented earlier, this section provides empirical evidence to assess its performance in real-world innovation scenarios. To evaluate the practical utility of the proposed TCI, we empirically examine its relationship with various indicators of patent quality. This evaluation serves two key purposes. First, it tests whether TCI meaningfully captures characteristics associated with higher-quality innovation outcomes. Second, it provides initial validation for the index's utility in applied research and innovation management contexts. The following sections report the results of these empirical studies in detail.

We select *First Claims* and *Forward Citations* as proxy indicators of patent quality based on their established theoretical foundations and empirical validation in innovation research (Marco et al., 2019; Moser et al., 2018). The *First Claim* represents the core inventive element of a patent and is often used to assess innovation and novelty (Allison et al., 2010; Mann & Underweiser, 2012). In contrast, the number of *Forward Citations* a patent receives is a well-established proxy for technological impact and recognition, reflecting the degree to which subsequent innovations draw upon the patented knowledge (Sun & Wright, 2022). According to prior research, these two indicators have been consistently validated as reliable measures of patent quality across diverse technological domains (Squicciarini et al., 2013).

### 4.3.1 Results of correlation analysis

To evaluate whether TCI meaningfully reflects broader innovation characteristics, we performed correlation analyses between TCI and patent quality. As shown in **Fig. 6** and **Fig. 7**, TCI generally displays a positive correlation with these indicators, both over time and across most IPC sections.

**Fig. 6** illustrates temporal correlations between TCI and patent quality across different periods, revealing consistently positive relationships with increasing strength over time. This progressive strengthening suggests that higher TCI values reliably correspond to improved patent technological quality and impact. **Fig. 7** presents IPC-sectional correlations, highlighting substantial variation across technological domains, with certain sectors exhibiting notably stronger positive correlations. Moreover, the correlation strength has grown more pronounced in recent years, implying that technological convergence is increasingly important in shaping innovation. These sectoral insights further validate TCI's practical relevance and underscore its utility for developing targeted innovation strategies tailored to specific technological domains.

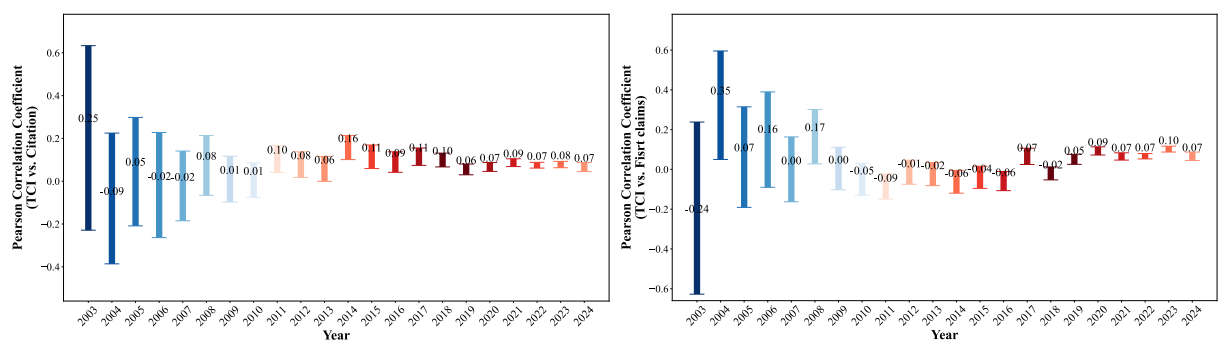

Fig.6a. Correlation between TCI and First Claims    Fig.6b. Correlation between TCI and Citation

**Fig. 6. Temporal correlations between TCI and patent quality**

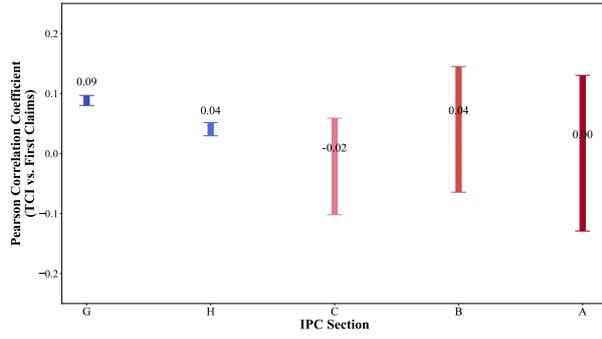 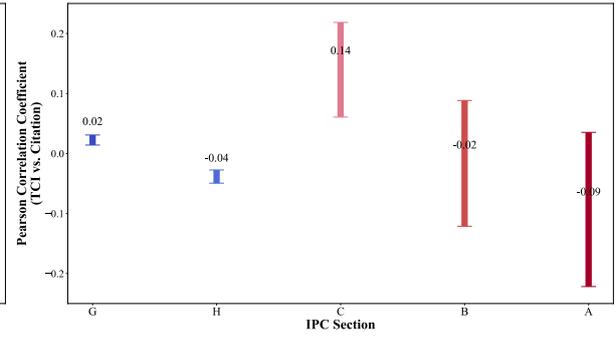

Fig.7a. Correlation between TCI and First Claims | Fig.7b. Correlation between TCI and Citation

**Fig.7 IPC-sectional correlations between TCI and patent quality**

### *4.3.2 Results of regression analysis*

To systematically evaluate the predictive validity of the TCI, we employed ordinary least squares (OLS) regression models in our empirical analysis. This methodological choice is motivated by OLS's interpretive clarity and its suitability for examining linear relationships, consistent with the theoretical framework that posits a linear association between technological convergence and various dimensions of patent quality (Ze-Lei et al., 2017). Prior research suggests that technological convergence enhances the breadth of knowledge integration (Borés et al., 2003; Park, 2017), promotes collaborative innovation, fosters greater novelty in technological development, and ultimately increases the technological impact of resulting patents (Caviggioli, 2016; Liu et al., 2020). Drawing on this theoretical foundation, we propose the following hypotheses to test the effect of TCI on patent quality:

**H1:** Technological convergence increases innovation novelty, as reflected by broader in scope first claims.

**H2:** Technological convergence enhances technological impact and recognition, as reflected by higher forward citation.

To isolate the effects of technological convergence, we incorporated several control variables: patent pages (*Pages*), patent claims (*Claims*), and backward citations (*Bcite*) for patent-specific characteristics that potentially influence outcomes, while year fixed effects (YearFE) control for unobserved macro-level factors including economic cycles, policy changes, and evolving technological landscapes. The regression model is thus specified as:

$$Y_{it} = \alpha + \beta TCI_{it} + \gamma \text{Controls}_{it} + YearFE + \varepsilon_{it} \qquad (15)$$

where $Y_{it}$ represents one of the patent quality and $Controls_{it}$ includes *Pages*, *Claims*, *Bcite*, and *Year* fixed effect. The inclusion of year fixed effects aims to control for unobserved macro-level changes, such as shifts in policy frameworks, economic conditions, or technological paradigms, that might otherwise bias the estimated relationship between TCI and patent characteristics indicators.

**Table. 2** summarizes the regression results, which consistently demonstrate statistically significant positive relationships between TCI and patent quality.

Specifically, in **Table. 2**, the analysis reveals a significant and positive correlation between TCI and *First Claims*, indicating that patents characterized by higher technological convergence demonstrate greater innovative novelty and broader technical scope. Furthermore, the positive correlation between TCI and *Forward Citations* substantiates that patents with higher convergence achieve greater recognition and exert more substantial influence within their respective technological domains. Across all models, the control variables generally conform to theoretical expectations. For instance, *Bcite* exhibits variable effects that appear contingent upon knowledge depth and patent scope considerations. Additionally, the incorporation of year fixed effects accounts for temporal variations in policy and economic environments, thereby enhancing the robustness of causal inferences in our analysis.

Table. 2 Results of regression analysis

| Variable | (1) First Claims | (2) Forward Citation |
|---|---|---|
| **Independent variable** | | |
| TCI | 0.009*** | 0.005*** |
| | (0.001) | (0) |
| **Control Variable** | | |
| Pages | 0.23*** | 0.082*** |
| | (0.006) | (0.004) |
| Claims | -0.292*** | 0.016*** |
| | (0.005) | (0.003) |
| Bcite | 0.017*** | 0.013*** |
| | (0.001) | (0.001) |
| _cons | -0.696*** | 3.614*** |
| | (0.045) | (0.027) |
| **fixed-effect** | | |
| Year | Yes | Yes |
| Observations | 87795 | 87795 |
| R-squared | 0.048 | 0.178 |

Overall, the regression results validate the practical effectiveness of the proposed TCI, confirming its robustness and utility in capturing real-world technological convergence patterns and innovation outcomes. Although the reported R² values appear relatively low, this outcome is both theoretically and empirically grounded (Brown et al., 1999; McFadden, 1972). As patent-level data inherently exhibit high heterogeneity, stemming from diverse organizational strategies, inventor characteristics, and technological uncertainties, substantial unexplained variance is expected. Furthermore, our dataset comprises over 87,000 patents, with skewed distributions and modest mean values across key variables. These attributes inherently limit the explanatory power of any single factor. Importantly, the consistency and statistical significance of coefficients across all four models underscore the robustness and validity of our findings, despite the modest R² values.

### 4.4 Robustness analysis

To establish the validity, robustness, and broad applicability of our novel TCI, we implemented a rigorous three-pronged validation approach encompassing alternative measurement methodologies, cross-version consistency analysis, and comparative regression testing.

Our research developed and evaluated seven distinct TCI variants (V1-V8), each representing different methodological approaches. These ranged from conventional IPC co-occurrence metrics such as Clustering Coefficient and Average Distance to advanced computational techniques, including SBERT, HGT, and integrated approaches that synthesize multiple information sources (HGT + Shannon, SBERT + Shannon, HGT + SBERT + Rao-Stirling). The culmination of this methodological progression is our proposed composite model (V8), which seamlessly integrates HGT, SBERT, and SDI principles. **Table. 3** presents a detailed comparison between baseline methods and our proposed approach.

**Table. 3 Comparison of Baseline and Proposed Methods for TCI**

| TCI Version | Method | Measurement |
|---|---|---|
| V1 | IPC Co-occurrence Models | Clustering Coefficient |
| V2 | IPC Co-occurrence Models | Average Distance |
| V3 | Transformer-based Models | SBERT |
| V4 | Graph-based Models | HGT |
| V5 | Composite Models | HGT + Shannon |
| V6 | Composite Models | SBERT + Shannon |
| V7 | Composite Models | HGT + SBERT + Rao-Stirling |
| V8 (Ours) | Composite Models | HGT + SBERT + Shannon |

*Note: V1-V7 represent baseline approaches derived from existing literature. V8 is the proposed composite method.*

The consistency assessment across these measurement approaches revealed both convergence and divergence patterns (see **Table 4**). While several traditional indices exhibited only modest correlations, the more advanced composite variants demonstrated markedly stronger alignment. Notably, V5 and V6 were highly correlated (0.904), reflecting their substantial methodological overlap. More importantly, our proposed index (V8) showed exceptionally strong associations with these leading benchmarks, 0.985 with V5 and 0.962 with V6, indicating that it successfully captures the common conceptual foundation of composite approaches while providing meaningful refinements. Spearman correlations reinforced these findings: V8 exhibited very high rank-order consistency with V5 (0.987), V6 (0.955), and V7 (0.925), underscoring its robustness across both absolute values and relative rankings. Together, these results provide compelling evidence of the convergent validity and methodological stability of our measure.

Regression analyses further highlighted the explanatory advantages of the proposed index (see **Table 5**). Across both dependent variables, first claims and forward citations, V8 consistently achieved the strongest or joint-strongest explanatory power. For first claims, it produced the highest coefficient (0.009, p<0.001) and an $R^2$ of 0.048, tying with V6 and outperforming other formulations. For citations, V8 again delivered the best performance (coef=0.005, p<0.001, $R^2$=0.178), matching V5 and V6 and exceeding all remaining versions. In contrast, earlier indices such as V1 and V2 exhibited clear limitations: V1 showed a negative relationship with first claims and only modest explanatory power for citations, while V2 failed to reach significance for first claims.

Taken together, the cross-method correlation and regression results demonstrate that our proposed TCI (V8) not only aligns closely with the most robust composite benchmarks but also offers superior explanatory strength. These findings confirm its theoretical soundness and practical value as a reliable measure of technological convergence in innovation research.

Table. 4 Correlation analysis among TCI versions

|  | V1 | V2 | V3 | V4 | V5 | V6 | V7 | V8 (Ours) |
|---|---|---|---|---|---|---|---|---|
| **Pearson_corr** | | | | | | | | |
| V1 | 1 | -0.044 | 0.385 | 0.403 | 0.4 | 0.458 | 0.273 | 0.434 |
| V2 | -0.044 | 1 | 0.178 | 0.15 | 0.027 | 0.084 | -0.051 | 0.048 |
| V3 | 0.385 | 0.178 | 1 | 0.964 | 0.562 | 0.753 | 0.247 | 0.644 |
| V4 | 0.403 | 0.15 | 0.964 | 1 | 0.703 | 0.806 | 0.454 | 0.754 |
| V5 | 0.4 | 0.027 | 0.562 | 0.703 | 1 | 0.904 | 0.863 | 0.985 |
| V6 | 0.458 | 0.084 | 0.753 | 0.806 | 0.904 | 1 | 0.618 | 0.962 |
| V7 | 0.273 | -0.051 | 0.247 | 0.454 | 0.863 | 0.618 | 1 | 0.789 |
| V8 (Ours) | 0.434 | 0.048 | 0.644 | 0.754 | 0.985 | 0.962 | 0.789 | 1 |
| **Spearman_corr** | | | | | | | | |
| V1 | 1 | 0.183 | 0.305 | 0.313 | 0.447 | 0.49 | 0.41 | 0.473 |
| V2 | 0.183 | 1 | 0.672 | 0.601 | 0.54 | 0.603 | 0.483 | 0.563 |
| V3 | 0.305 | 0.672 | 1 | 0.637 | 0.245 | 0.442 | 0.187 | 0.305 |
| V4 | 0.313 | 0.601 | 0.637 | 1 | 0.495 | 0.418 | 0.445 | 0.467 |
| V5 | 0.447 | 0.54 | 0.245 | 0.495 | 1 | 0.908 | 0.954 | 0.987 |
| V6 | 0.49 | 0.603 | 0.442 | 0.418 | 0.908 | 1 | 0.816 | 0.955 |
| V7 | 0.41 | 0.483 | 0.187 | 0.445 | 0.954 | 0.816 | 1 | 0.925 |
| V8 (Ours) | 0.473 | 0.563 | 0.305 | 0.467 | 0.987 | 0.955 | 0.925 | 1 |

Table. 5 Regression Analysis of Baseline vs. Proposed TCI Measures

| Y Variable | TCI Version | Coef | P-Value | R2 |
|---|---|---|---|---|
| First Claims | **V8 (Ours)** | **0.009** | **0** | **0.048** |
| | V6 | 0.009 | 0 | 0.048 |
| | V5 | 0.008 | 0 | 0.048 |
| | V7 | 0.004 | 0 | 0.048 |
| | V4 | 0.001 | 0 | 0.046 |
| | V3 | 0.001 | 0 | 0.045 |
| | V2 | 0 | 0.907 | 0.045 |
| | V1 | -0.004 | 0 | 0.046 |
| Citation | **V8 (Ours)** | **0.005** | **0** | **0.178** |
| | V6 | 0.005 | 0 | 0.178 |
| | V5 | 0.005 | 0 | 0.178 |
| | V1 | 0.004 | 0 | 0.177 |
| | V2 | 0.003 | 0.008 | 0.176 |
| | V7 | 0.002 | 0 | 0.177 |
| | V4 | 0.001 | 0 | 0.177 |
| | V3 | 0.001 | 0 | 0.176 |

# 5. Discussion and conclusions

## 5.1 Key findings

Measuring technological convergence remains a persistent challenge due to its multidimensionality and complexity. Using empirical data from Chinese twin transition patents from 2003 to 2024, our analysis identifies distinct evolutionary trends in technological convergence. Significant variability in convergence intensity emerges across IPC sections, with Physics and Electricity exhibiting pronounced multimodal distributions. In contrast, sections such as Human Necessities, Performing Operations, and Chemistry present relatively balanced and evenly distributed multi-technology convergence.

Our regression analyses confirm statistically significant, positive correlations between TCI and patent quality. Higher TCI values notably correlate with increased innovation novelty (as measured by *First Claims*) and enhanced technological impact (demonstrated through *Forward Citations*). These empirical findings robustly support our theoretical framework, demonstrating that patents characterized by higher technological convergence consistently facilitate deeper knowledge integration and produce more influential innovations within their technological fields.

Furthermore, our comprehensive robustness evaluation comparing TCI against six alternative convergence metrics confirms that our method consistently achieves superior correlation strength and explanatory power across multiple validation tests, establishing its methodological stability and measurement reliability.

## 5.2 Contributions to theory

Our main theoretical contribution lies in developing a comprehensive framework for measuring technological convergence that simultaneously incorporates both cross-domain knowledge depth and technological portfolio breadth. Drawing from the structural attributes of patent text, we propose TCI, a novel approach that integrates HGT, SBERT, and SDI methodologies to enable the analysis of technological convergence in a more granular and operationalizable manner. This conceptualization moves beyond traditional proxies and offers an enriched understanding of how convergence unfolds across multiple knowledge spaces.

We further establish a "measurement-utility" feedback loop that connects the TCI to innovation outcomes. Through empirical validation, we demonstrate that technologies with higher TCI values not only reflect greater cross-domain integration but also predict stronger patent quality. This validation provides a methodological advancement, ensuring that convergence metrics are both theoretically grounded and practically relevant.

Our findings suggest that technological convergence should be understood not only as a structural phenomenon but also as an important predictor of innovation outcomes (Lee et al., 2018). Organizations aiming to leverage convergence for competitive advantage should consider strategies that foster both deeper knowledge integration and broader technological diversification (Lin & Chen, 2008). Similarly, policymakers could design adaptive innovation policies that reflect the layered complexity of technological evolution, rather than relying on uniform support measures across industries (Dolata, 2009).

Finally, the flexible architecture of the TCI offers potential for broader application across different analytical levels and industrial contexts. By enabling researchers and practitioners to trace convergence patterns at the patent, firm, and sectoral levels, our framework supports a more comprehensive exploration of how technological convergence shapes innovation trajectories in an increasingly interconnected landscape.

**5.3 Implications for practice**

Our research on TCI offers several practical implications for innovation stakeholders that enable more informed decision-making. For entrepreneurs, TCI highlights emerging technological fields where deeper knowledge integration enhances the defensibility of intellectual property (Roco et al., 2013), while broader technological diversity expands scalable market opportunities (Muldoon et al., 2023). Depth values exceeding the sector median indicate domains where original core-periphery combinations remain under-exploited, as reflected by higher first-claim significance. Meanwhile, high breadth flags heterogeneous application possibilities, as evidenced by greater forward citation impact. Using both dimensions to balance exploratory and exploitative investments improves portfolio resilience and aligns early-stage capital with longer-term growth prospects (He et al., 2022).

Inventors and R&D managers can integrate TCI thresholds into project-evaluation frameworks. Projects occupying the upper-right area of the convergence landscape, characterized by strong knowledge integration and high portfolio diversity, exhibit a stronger propensity for breakthrough claims and downstream impact (Cho et al., 2015). This dual-dimensional assessment helps research teams avoid excessive specialization while maintaining focus, thereby optimizing resource allocation toward high-potential technological trajectories.

Policymakers and funding agencies may adopt TCI as an evidence-based criterion for allocating grants, tax incentives, and procurement contracts (Georghiou et al., 2014). Because the index outperforms six established benchmarks in predicting patent quality, it provides a more detailed basis for targeting resources toward technologies that promise high social spill-overs. Incorporating depth-breadth targets into evaluation frameworks can shift incentives away from superficial patent accumulation toward substantive knowledge integration.

In addition, twin transition challenges are particularly well-addressed by TCI's integrative approach (Myshko et al., 2024). In analyzing digital-sustainability convergence patterns, our index reveals where specialized knowledge domains remain disconnected despite potential synergies. The bimodal distribution in Physics and Electricity sections, for instance, points to specific bridging opportunities between deep-tech power electronics specialists and broader clean-energy applications. By targeting these convergence gaps, stakeholders can develop more effective technological solutions that simultaneously address digitalization and decarbonization imperatives.

**5.4 Limitation and future research**

We also acknowledge certain limitations in this study. Although our analysis provides meaningful insights into technological convergence patterns, it is bounded by specific scope conditions. First, the empirical evidence is derived from Chinese twin transition patents, which may limit external validity. Future research could extend the application of TCI to patent and non-patent datasets from other major innovation systems, such as Europe, the United States, Japan, and South Korea, to more fully assess its cross-cultural applicability. Second, while our regression analyses establish strong predictive associations, a more comprehensive causal understanding could be pursued by leveraging exogenous policy shocks as quasi-natural experiments. Third, although we examine TCI within the context of the twin transition, broader applications across emerging interdisciplinary domains offer promising directions for future research.